\documentclass[pdflatex,sn-nature]{sn-jnl}%

\usepackage{graphicx}%
\usepackage{multirow}%
\usepackage{amsmath,amssymb,amsfonts}%
\usepackage{amsthm}%
\usepackage{mathrsfs}%
\usepackage[title]{appendix}%
\usepackage{xcolor}%
\usepackage{textcomp}%
\usepackage{manyfoot}%
\usepackage{booktabs}%
\usepackage{algorithm}%
\usepackage{algorithmicx}%
\usepackage{algpseudocode}%
\usepackage{listings}%
\usepackage{lineno}

\theoremstyle{thmstyleone}%
\theoremstyle{thmstyletwo}%

\theoremstyle{thmstylethree}%

\begin{document}

\title[Article Title]{DiffQEC: A versatile diffusion model for quantum error correction}

\author[1,2]{\fnm{Tianyi} \sur{Xu}}
\author[1,2]{\fnm{Qinglong} \sur{Liu}}
\author[1]{\fnm{Maolin} \sur{Wang}}
\author[1]{\fnm{Fei} \sur{Zhang}}

\author[1]{\fnm{Zhe} \sur{Zhao}}
\author*[3]{\fnm{Yang} \sur{Wang}}\email{ywang.qt@gmail.com}
\author*[1,2]{\fnm{Ye} \sur{Wei}}\email{ye.wei@cityu.edu.hk}

\affil[1]{\orgdiv{Department of Data Science}, \orgname{City University of Hong Kong}, \orgaddress{ \country{China}}} 
\affil[2]{\orgdiv{Department of Materials Science}, \orgname{City University of Hong Kong}, \orgaddress{\country{China}}} 
\affil[3]{\orgdiv{3rd Institute of Physics}, \orgname{University of Stuttgart}, \orgaddress{\street{Allmandring 13}, \city{Stuttgart}, \postcode{70569}, \country{Germany}}}


\abstract{
Quantum computers could solve problems beyond the reach of classical devices, but this potential depends on quantum error correction (QEC) to protect fragile quantum states from noise. A central challenge in QEC is decoding: inferring likely physical errors from syndrome patterns generated by repeated stabilizer measurements. Existing decoders, including graph-based and neural approaches, typically return a single correction hypothesis and therefore discard the richer posterior structure of the error distribution conditioned on the observed syndrome. Here we recast QEC decoding as posterior inference using discrete denoising diffusion, exploiting the analogy between stochastic error accumulation and the forward diffusion process. We introduce DiffQEC, a generative decoder that combines a syndrome processor for multi-round spatial-temporal syndrome histories with syndrome feature modulation to condition denoising on the observed syndrome throughout inference. On experimental data from Google's superconducting quantum processor, DiffQEC reduces logical error rates by up to 10.2\% relative to minimum-weight perfect matching and by about 5\% relative to tensor-network decoding. These improvements persist for larger code distances up to 17 under depolarizing noise and for logical circuits of increasing depth. Beyond accuracy, the learned posterior provides confidence estimates for post-selection and reveals physically meaningful error structure, establishing posterior generative decoding as a practical framework for QEC.
}

%


\maketitle
\section*{Introduction}
\label{sec1}

Quantum computing could enable solutions to problems beyond the reach of classical machines, but this promise depends on quantum error correction (QEC) to protect fragile quantum information from noise~\cite{Acharya2024,Wang2024SA,Gidney2025,Zhou2025,Bravyi2024Nature}. In QEC, repeated stabilizer measurements produce spatiotemporal syndrome patterns that record how physical errors arise, propagate and accumulate during computation~\cite{Terhal2015,Wang2023PhdThesis,Bausch2024}. Decoding these syndromes into effective corrections is therefore a central challenge for practical fault-tolerant quantum computing.

Many existing decoders either rely on explicit noise models that may not fully capture real hardware behaviour, or learn direct discriminative mappings from syndrome to correction, which provide limited access to the underlying uncertainty structure~\cite{deMartiiOlius2024quantum,Higgott2021,Wu2025,Piveteau2024,Bausch2024,Varbanov2025,Zhou2025NCS}.
Yet most ultimately reduce inference to a single correction hypothesis. Even approaches that estimate marginal error probabilities, such as belief propagation~\cite{deMartiiOlius2024quantum,Roffe2020PRR}, generally do not maintain a tractable representation of the full posterior over error configurations. This discards uncertainty information that could support confidence estimation, selective decision-making and downstream analysis of the underlying noise mechanisms.
The limitation becomes more acute in the large-code, long-circuit regime. To maintain bounded memory and low latency, scalable implementations often partition syndrome histories into sequential temporal windows~\cite{Skoric2023,Tan2023PRXQ,Serra2025PRXQ,Turner2026PRXQ}. In such settings, information is typically passed across boundaries through hard correction decisions rather than quantified uncertainty, obscuring long-range correlations induced by logical operations and correlated device-specific noise~\cite{Turner2026PRXQ,Serra2025PRXQ,Cain2024,Nickerson2019,Bausch2024,Acharya2024,Tiurev2023Quantum}. These considerations motivate decoding as posterior inference over complex error distributions conditioned on syndrome measurements.

\begin{figure}[h]
\centering
\includegraphics[width=0.98\textwidth]{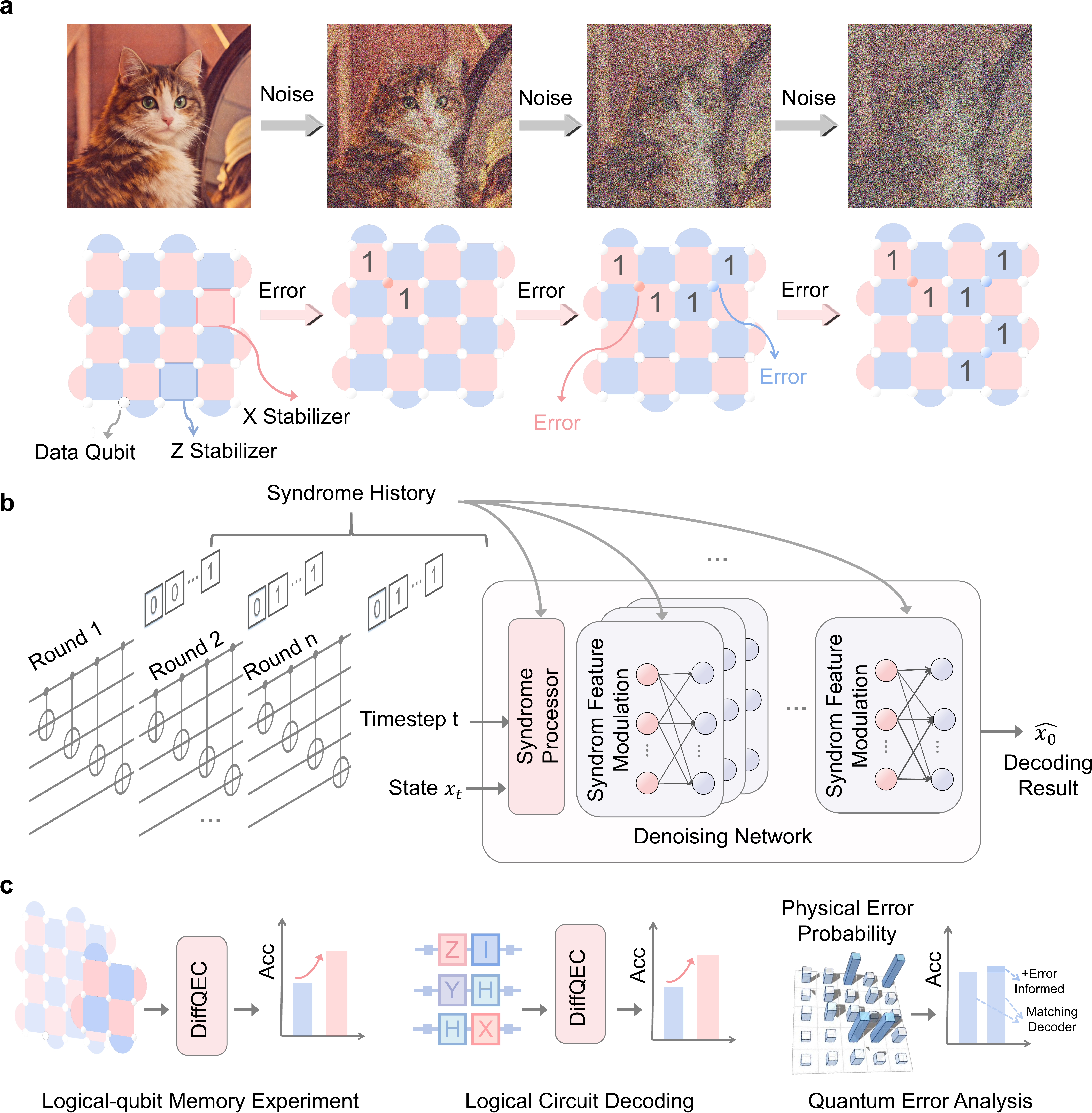}
\caption{Overview of DiffQEC. 
(a) Illustration of the analogy between the forward denoising diffusion process and the progressive accumulation of errors in the surface code. 
(b) The decoding pipeline of DiffQEC. Multi-round syndrome histories are encoded and used to condition the denoising network together with the diffusion timestep $t$ and current noisy state $x_t$, producing the estimated logical correction $\hat{x}_0$. 
(c) Application scenarios of DiffQEC, including logical-qubit memory decoding, logical circuit decoding, and exploratory quantum error analysis.
}\label{fig1}
\end{figure}

Denoising diffusion probabilistic models~\cite{ho2020denoising,austin2021structured,pmlr-v37-sohl-dickstein15} provide a natural framework for this view. The accumulation of physical errors in quantum devices mirrors the forward diffusion process, in which information is progressively corrupted over time. As illustrated in Fig.~\ref{fig1}a, this analogy suggests decoding as a reverse denoising process conditioned on syndrome measurements, enabling iterative refinement of candidate error configurations while capturing spatial and temporal correlations within a unified framework.

Here, we introduce DiffQEC, a diffusion-based decoder that infers corrections from measured syndromes through iterative denoising. As summarized in Fig.~\ref{fig:fig2}, DiffQEC integrates a syndrome processor for encoding multi-round syndrome histories with a syndrome feature modulation module that conditions the denoising network throughout inference. This architecture enables the model to capture structured spatiotemporal error correlations while progressively refining corrections in the discrete domain. DiffQEC outperforms point-estimation baselines on Google’s experimental quantum-memory data for distance-3 and distance-5 surface codes, with gains that increase with code size, and retains this advantage up to distance~17 in simulated memory experiments and across deeper logical circuits. The learned posterior further supports calibrated post-selection and reveals physically meaningful error structure through attribution analysis, establishing diffusion-based posterior inference as a practical generative framework for QEC decoding.

\section*{Results}

\subsection*{DiffQEC provides a general generative framework for decoding}
\begin{figure}[h]
\centering
\includegraphics[width=1\textwidth]{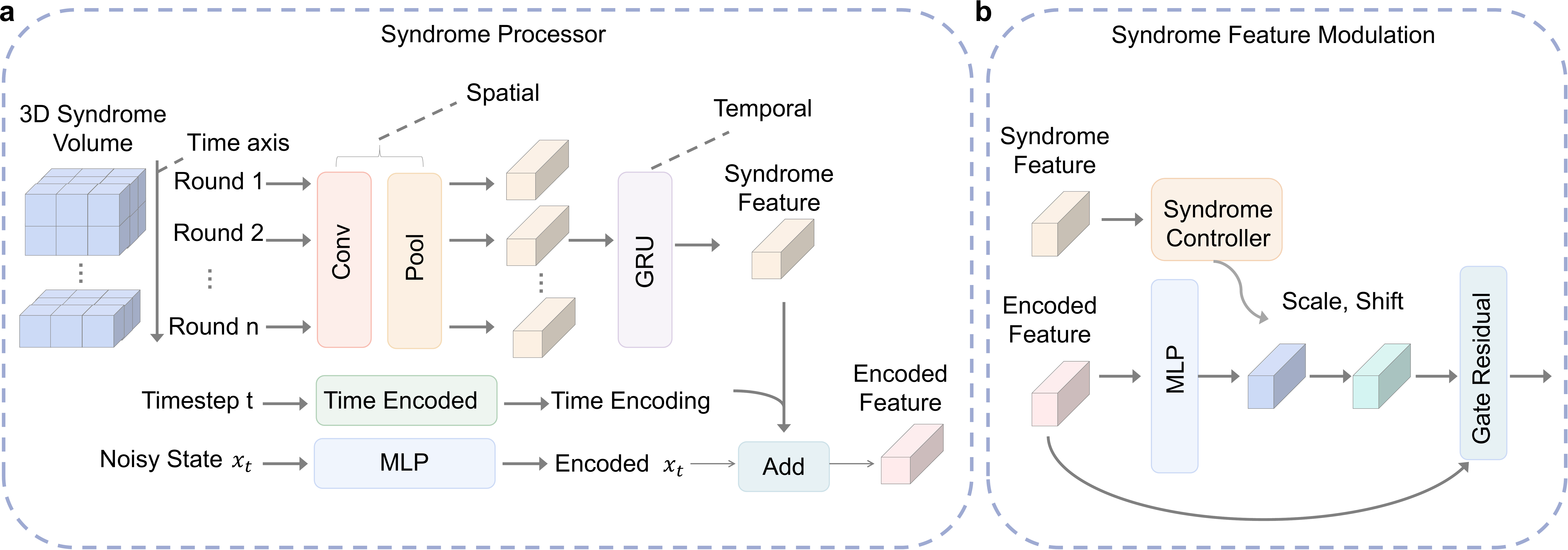}
\caption{Architecture of DiffQEC. 
(a) Syndrome processor. 
A 3D syndrome volume is encoded through per-round spatial convolution and pooling, followed by temporal aggregation across rounds with a GRU to produce a syndrome feature. In parallel, the diffusion timestep $t$ and the current noisy state $x_t$ are encoded separately. These representations are combined with the syndrome feature to form the encoded feature for denoising. 
(b) Syndrome feature modulation. The encoded feature is further conditioned by the syndrome feature through a syndrome controller, which generates scale and shift signals together with a gate residual pathway before producing bitwise logits in the denoising network.}\label{fig:fig2}
\end{figure}
Before presenting the quantitative results, we first summarize the design rationale of DiffQEC (Fig.~\ref{fig:fig2}). The architecture is motivated by two features of practical QEC decoding. First, multi-round syndrome histories contain structured correlations across both space and time, so an effective decoder should capture not only local spatial patterns within each round but also temporal dependencies across repeated rounds. Second, decoding is not merely a static mapping from syndrome to correction: under realistic noise, the same intermediate correction hypothesis may require different refinements depending on the observed syndrome context. These considerations motivate an architecture that combines spatiotemporal syndrome encoding with syndrome-conditioned iterative denoising.

The syndrome processor therefore arranges the multi-round syndrome history as a 3D input volume stacked along the temporal dimension. Spatial features are extracted within each round using convolution and pooling, and are then aggregated across rounds by a gated recurrent unit (GRU)~\cite{cho-etal-2014-learning} to produce a compact syndrome representation. In parallel, the diffusion timestep $t$ and the current noisy state $x_t$ are encoded separately, and their embeddings are integrated with the syndrome representation to form the conditioning feature used for denoising. This design enables DiffQEC to represent spatiotemporal syndrome structure while iteratively refining candidate corrections in the discrete domain.
As shown in Fig.~\ref{fig:fig2}b, conditioning is applied throughout the denoising network rather than only at the input layer. A syndrome controller generates feature-wise modulation signals, together with a gated residual pathway, so that denoising remains responsive to the observed syndrome at each stage of inference. This enables the model to adapt its refinement trajectory to the measurement context rather than relying on a fixed update rule.

Conceptually, this iterative inference procedure is related to message-passing decoders such as belief propagation (BP)~\cite{deMartiiOlius2024quantum,Roffe2020PRR}. The distinction is that BP derives its updates explicitly from the bipartite graph connecting data qubits and stabilizer checks, whereas DiffQEC learns denoising dynamics directly from data. This generative formulation also accommodates the degeneracy of stabilizer codes, in which a single syndrome may correspond to an equivalence class of underlying error configurations.

\begin{figure}[t]
\centering
\includegraphics[width=1\textwidth]{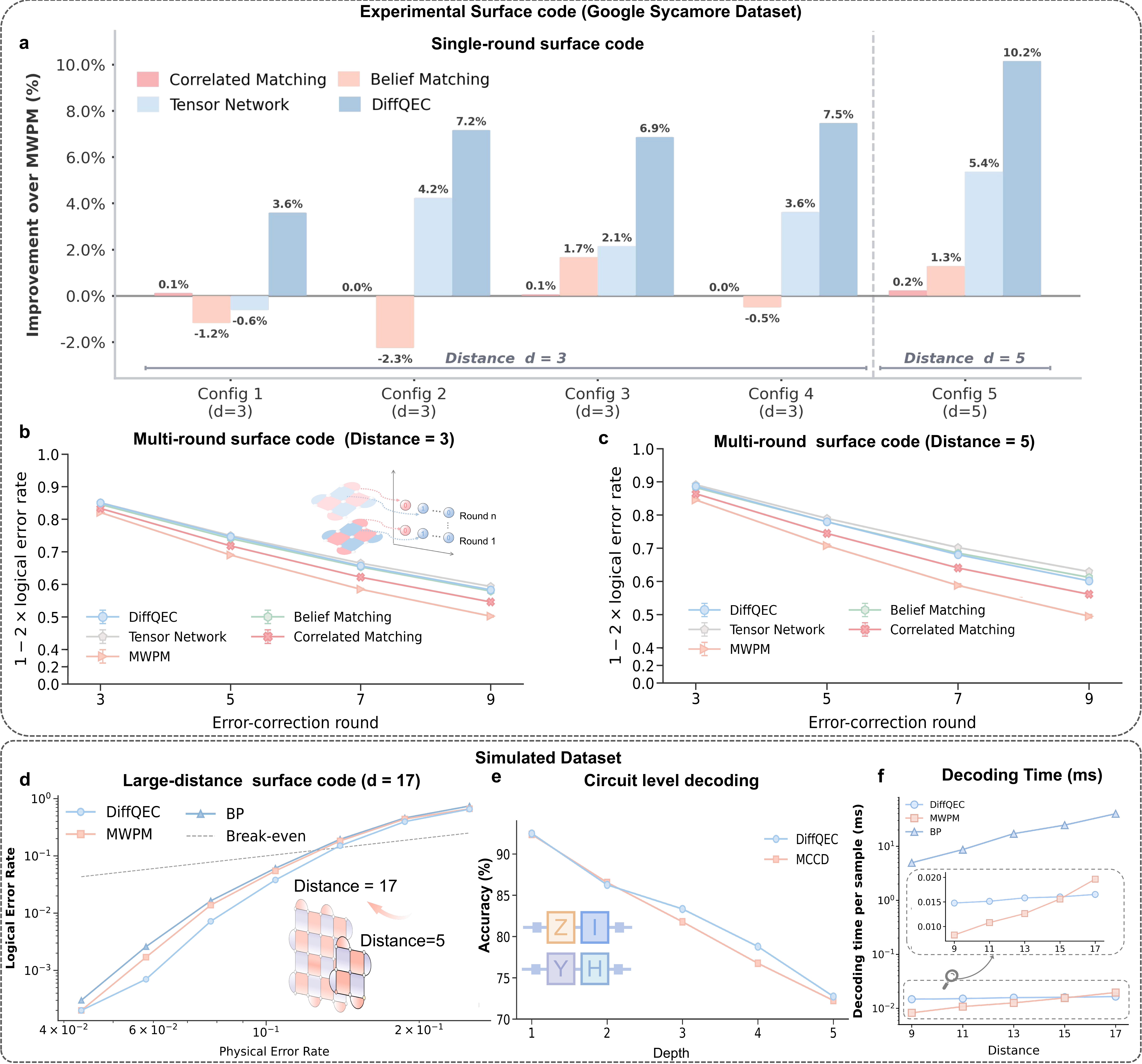}
\caption{Performance of DiffQEC on Google’s Sycamore surface-code, simulated multi-round surface-code, large-distance surface code and circuit-level decoding. (a) Relative improvements of over MWPM on Google's Sycamore memory dataset. (b) and (c) show multi-round simulated surface-code decoding results at distances $3$ and $5$. (d) Large-distance decoding performance at distance $17$ as a function of physical error rate. 
(e) Decoding time of DiffQEC and MWPM across increasing code distance.
(f) Circuit-level decoding accuracy compared with MCCD .}\label{fig:result1}
\end{figure}

\begin{figure}[t]
\centering
\includegraphics[width=0.99\textwidth]{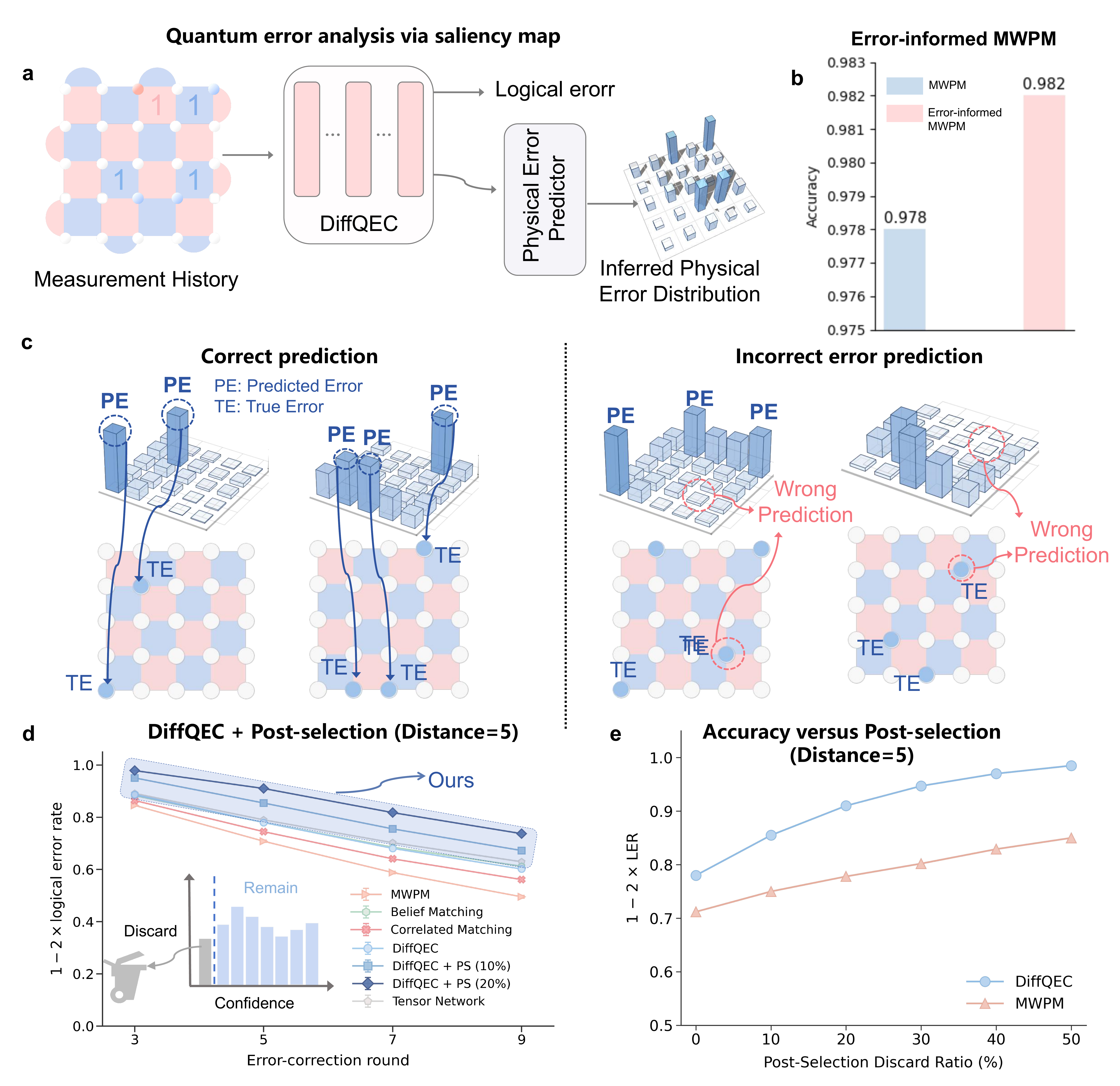}
\caption{Quantum error analysis, inference speed, and confidence-based post-selection with DiffQEC. (a) Measurement history is processed by the trained decoder to derive an attribution-based score over candidate physical error locations. (b) Incorporating this inferred error information into MWPM yields an improvement in decoding accuracy. (c) Representative visualizations of the inferred physical-error maps, including cases where the dominant predicted support agrees with the true error pattern and a representative failure case. Here, PE and TE denote predicted error and true error, respectively. 
 (d) Multi-round decoding performance at distance 5 with post-selection (PS), where PS denotes confidence-based post-selection obtained by discarding a fraction of the least-confident samples. Results are shown for DiffQEC without post-selection and with discard rates of 10\% and 20\%, together with baseline decoders. (e) Decoding performance versus post-selection discard ratio at distance 5 for DiffQEC and MWPM.
}\label{fig:fig4}
\end{figure}

\subsection*{Performance on Google’s Sycamore experimental data}
We first evaluated DiffQEC on Google’s publicly released dataset from Sycamore surface-code memory experiments~\cite{Bausch2024}, which provides experimentally collected stabilizer measurement records for rotated surface codes with distances $d=3$ and $d=5$. This dataset offers a stringent benchmark under realistic hardware noise, beyond simulations based on simplified error models. In this memory-experiment setting, the decoding task is to predict whether a logical error has occurred on each logical qubit.

We formulate this task as learning the conditional distribution $p_\theta(l \mid s)$, where $s$ denotes the measured syndrome and $l \in \{0,1\}^{L}$ is a binary label vector indicating the presence or absence of logical errors on the corresponding logical qubits. Consistent with this formulation, DiffQEC operates directly on binary syndrome histories, preserving the discrete structure of stabilizer measurements throughout inference. This allows the model to capture structured correlations induced by realistic hardware noise when estimating logical-error probabilities.

The single-round results for five hardware configurations are summarized in Fig.~\ref{fig:result1}a, which reports the relative reduction in logical error rate with respect to minimum-weight perfect matching (MWPM), a standard graph-based decoding baseline. DiffQEC consistently improves on MWPM across all configurations. For the distance-$3$ codes, the relative reduction reaches up to $7.5\%$, while for the distance-$5$ code it increases to $10.2\%$, suggesting that the advantage of posterior-based decoding becomes more pronounced in more complex decoding regimes. We also compare with tensor-network (TN) decoding, a strong high-accuracy reference method for the same Sycamore memory setting. TN achieves strong performance by evaluating many error configurations consistent with the observed detection events, but at substantially higher computational cost. In the single-round setting, DiffQEC attains lower logical error than TN.

We next assessed whether DiffQEC can exploit temporal structure across repeated syndrome measurements by evaluating it in the multi-round decoding setting, where syndrome information accumulates over multiple measurement cycles. In this regime, the decoder must integrate temporal dependencies rather than rely on a single-round observation. As shown in Fig.~\ref{fig:result1}b,c, DiffQEC maintains high decoding accuracy for both distance-$3$ and distance-$5$ codes, achieving performance comparable to TN across the evaluated multi-round settings. For distance-$3$, DiffQEC already matches TN under repeated syndrome measurements, and this comparable performance is retained at distance-$5$, where the syndrome history spans a richer space of temporal error patterns.

This accuracy is accompanied by a substantially more favorable computational profile. Even in the single-round distance-$5$ setting, TN requires about $0.146$ ms per sample, reflecting the cost of evaluating many candidate error configurations. By contrast, DiffQEC requires inference time on the order of $10^{-2}$ ms per sample in our measurements, while maintaining comparable decoding accuracy. These results show that DiffQEC combines strong empirical performance on experimental hardware data with efficient inference under both single-round and multi-round decoding.

\subsection*{Superior performance at larger code distances and in circuit-level decoding}
Having established strong performance on experimental hardware data, we next examined DiffQEC in simulated settings to assess its scalability in more demanding regimes, including larger code distances and logical circuits beyond memory experiments. These controlled settings provide access to larger inputs and richer spatiotemporal correlations, making them well suited for evaluating decoder scalability.
We considered two complementary dimensions of scaling: increasing the code distance in memory decoding and extending to logical-circuit decoding. In logical circuits, stabilizer measurements are interleaved with logical gate operations, which can propagate errors across qubits and over time, giving rise to more complex spatiotemporal correlations than in memory experiments. 

In the memory setting, DiffQEC maintains strong decoding performance as the surface-code distance increases up to 17 and generally outperforms MWPM across most of the tested physical error range (Fig.~\ref{fig:result1}d). In representative low-to-intermediate error regimes, the logical error rate is approximately \(7\times 10^{-4}\) for DiffQEC, compared with \(1.7\times 10^{-3}\) for MWPM, whereas in a higher-error regime the corresponding values are about \(4\times 10^{-2}\) and \(5\times 10^{-2}\), respectively.

We next evaluated circuit-level decoding by comparing DiffQEC with MCCD~\cite{Zhou2025NCS}, an LSTM-based decoder developed specifically for logical-circuit decoding with dedicated modules for gate operations. We considered logical circuit depths from 1 to 5, where depth denotes the number of logical gate layers. DiffQEC remains close to MCCD at shallow depths and shows a clearer advantage as circuit depth increases. As shown in Fig.~\ref{fig:result1}f, DiffQEC achieves 92.35\% accuracy at depth 1 and 72.26\% at depth 5, compared with 91.71\% and 69.64\% for MCCD, respectively. Considering that DiffQEC was not specifically designed for logical-circuit decoding, these results indicate the potential of diffusion-based decoding in more structured circuit-level settings.

In addition to accuracy, decoding latency is a key practical measure of scalability. In fault-tolerant quantum computing, non-Clifford operations essential for universal computation typically require real-time decoding of syndrome histories together with the corresponding corrections before computation can proceed. Excessive latency would lead to an accumulation of unprocessed syndrome data and hinder scalable operation~\cite{Skoric2023}. As shown in Fig.~\ref{fig:result1}e, the latency of DiffQEC in the memory setting remains broadly comparable to that of MWPM. Its runtime increases more gradually with code distance and remains within the millisecond regime, consistent with the requirements of real-time decoding in superconducting hardware~\cite{Bausch2024}.

\subsection*{DiffQEC learns posterior structure beyond logical prediction}
Having established DiffQEC as an accurate and scalable decoder, we next asked whether its probabilistic outputs contain useful information beyond the final logical prediction. In particular, we examined whether these outputs provide informative confidence estimates for post-selection and whether they reflect meaningful structure in the conditional distribution of physical errors given the observed syndrome.

We first tested whether the probabilistic outputs of DiffQEC provide useful confidence information. To this end, we used the predicted logical-error probabilities as confidence scores and performed post-selection by discarding a fixed fraction of the least confident samples. As shown in Fig.~\ref{fig:fig4}d, this consistently improves performance at distance~5 for discard ratios of 10\% and 20\%. Fig.~\ref{fig:fig4}e further shows that \(1 - 2 \times \mathrm{LER}\) increases monotonically as the discard ratio grows from 0\% to 50\% for both DiffQEC and MWPM. Across all settings, DiffQEC yields systematically better sample ranking, indicating that its probabilistic outputs provide more informative confidence estimates for post-selection.

Next, we investigated whether the model also captures information related to the underlying physical errors. To  do so, we constructed saliency maps over physical qubits by first computing syndrome-level attribution scores using integrated gradients~\cite{sundararajan2017axiomatic} and then mapping these scores to physical qubits through the parity-check matrix. The resulting maps highlight candidate physical error locations (Fig.~\ref{fig:fig4}a). In representative examples, these saliency patterns are broadly consistent with the true physical error locations (Fig.~\ref{fig:fig4}c). Although they do not reconstruct the full physical error configuration, they suggest that DiffQEC captures meaningful physical-error-related structure beyond the final binary logical decision.

Finally, we asked whether this inferred information could assist downstream decoding. As a simple exploratory test, we incorporated the attribution-derived saliency scores as auxiliary side information for MWPM on a 2,500-sample subset of the distance-3 Google Sycamore dataset~\cite{Bausch2024}. As shown in Fig.~\ref{fig:fig4}b, this increases decoding accuracy from 0.978 to 0.982. Although the gain is modest, MWPM already performs near ceiling on this subset, leaving limited room for improvement. The result nevertheless suggests that the saliency information extracted from DiffQEC contains useful cues for decoding, and that such cues may become more valuable in more challenging regimes with stronger spatial and temporal error correlations.

Taken together, these results indicate that DiffQEC learns posterior structure that extends beyond the final logical prediction. This is reflected both in informative confidence estimates that support effective post-selection and in physical-error-related cues that can be visualized and potentially exploited for downstream decoding.

\section*{Discussion}\label{sec13}

Our results suggest that diffusion-based generative decoding provides a practical alternative to purely discriminative syndrome-to-decision mappings in QEC. Rather than treating decoding only as a direct prediction problem, DiffQEC frames it as an iterative denoising process over discrete correction variables. This formulation appears to be well matched to QEC settings in which syndrome histories contain nontrivial spatial and temporal correlations, including repeated stabilizer measurements and circuit-level error propagation. More broadly, the present results indicate that discrete diffusion models can serve as flexible decoders across a range of decoding regimes without requiring task-specific redesign for each setting.


Beyond the final decoding decision, DiffQEC also exposes information that can be used in other ways. In particular, the model supports effective post-selection, with decoding performance on the retained set improving as increasingly uncertain samples are discarded. It also reveals physical-error-related cues that can be visualized through saliency maps and explored for downstream decoding. Although these signals do not amount to a full reconstruction of the underlying physical error pattern, they suggest that generative decoders may offer a richer interface to QEC than point prediction alone. In this sense, diffusion-based decoding may be valuable not only for logical-error suppression, but also for exposing uncertainty and error-related information that can be used in later stages of inference.

Beyond the settings evaluated here, the generative decoding framework points to several promising extensions. In the current multi-round implementation, a single model is jointly trained across syndrome histories of different temporal lengths, which is practically convenient and may encourage the sharing of temporal representations across related decoding tasks, although some loss of round-specific specialization may also arise. More fundamentally, the present framework targets the final logical correction directly. A natural next step is therefore to move toward richer modeling of the underlying noise process itself, especially in multi-round and circuit-level settings where errors accumulate, propagate, and correlate over time. 

One promising direction is to introduce latent physical-error hypotheses or uncertainty-aware intermediate states while retaining supervision at the logical level. Such representations could support window-based decoding schemes that pass forward structured uncertainty rather than only hard correction decisions, thereby better preserving correlations that extend across temporal boundaries. A direct way to realize this idea would be to use simulation-based supervision for physical-error targets. However, because such targets are generally unavailable in real hardware experiments and simulated noise may not faithfully capture device-specific error mechanisms, we do not view this as the preferred route. Instead, a more practical direction may be to continue training primarily on logical outcomes while introducing auxiliary objectives that enforce consistency with the observed syndrome history across neighboring windows and the structural constraints of the underlying circuit. 

Taken together, these directions suggest a path from the current general-purpose generative decoder toward one that models the noise process more explicitly and is better aligned with the requirements of scalable fault-tolerant computation.
\backmatter

\bmhead{Acknowledgements}
This work was supported by the German Federal Ministry of Education and Research (BMBF) through the project QECHQS with Grant No. 16KIS1590K.

\bmhead{Note added}
During the finalization of this manuscript, we became aware of an independent work exploring diffusion-based decoding for quantum low-density parity-check (qLDPC) codes \cite{Liu2025}.




\section*{Method}
\label{sec:method}

\subsection*{Discrete Conditional Diffusion Process}
To realize the conditional formulation, DiffQEC adopts a discrete denoising diffusion process that is naturally matched to the binary nature of the logical correction representation. A clean logical correction vector $x_0 \in \{0,1\}^{L}$ is progressively corrupted into $x_t$ over $T$ steps through a sequence of binary Markov kernels $\{Q_t\}_{t=1}^T$:
\[
q(x_t \mid x_{t-1}) = Q_t(x_t \mid x_{t-1}),
\qquad
q(x_{1:T}\mid x_0)=\prod_{s=1}^{T} q(x_s\mid x_{s-1}).
\]
The marginal transition from $x_0$ to $x_t$ is given by
\[
q(x_t\mid x_0)=\bar Q_t(x_t\mid x_0),
\qquad
\bar Q_t = Q_1 Q_2 \cdots Q_t,
\]
where $L$ denotes the dimension of the logical correction representation used by the decoder.

Each transition matrix $Q_t$ is chosen as a binary symmetric corruption kernel,
\[
Q_t \;=\;
\begin{bmatrix}
1 - \tfrac{\beta_t}{2} & \tfrac{\beta_t}{2} \\[3pt]
\tfrac{\beta_t}{2} & 1 - \tfrac{\beta_t}{2}
\end{bmatrix},
\]
where the noise schedule $\{\beta_t\}_{t=1}^T$ controls the strength of binary corruption at each diffusion step. 
We employ a cosine schedule~\cite{nichol2021improved} with offset $s=0.008$:
\[
\bar{\alpha}_t =
\frac{\cos^2\!\Big(\frac{\tfrac{t}{T}+s}{1+s}\,\frac{\pi}{2}\Big)}
     {\cos^2\!\Big(\frac{s}{1+s}\,\frac{\pi}{2}\Big)},
\qquad
\beta_t = 1 - \frac{\bar{\alpha}_t}{\bar{\alpha}_{t-1}},
\]
where $\bar{\alpha}_t$ denotes the cumulative signal-retention coefficient up to 
step $t$. The resulting betas are clipped to 
$[10^{-4},\,0.9999]$ for numerical stability.
Unlike continuous Gaussian diffusion, this discrete formulation preserves the binary support of the logical variables throughout the forward process, making it directly compatible with the $\{0,1\}^{L}$ domain of the logical decoding task.

\subsection*{Reverse Diffusion and Decoding}

Decoding proceeds by reversing the discrete diffusion chain under the measured syndrome condition. Starting from an initial random state$x_T \sim \mathrm{Bernoulli}\!\left(\tfrac12\right)^L$,

DiffQEC iteratively refines the binary configuration through a sequence of reverse denoising steps,
\[
x_T \;\to\; x_{T-1} \;\to\; \cdots \;\to\; x_1 \;\to\; x_0.
\]

The corresponding denoising architecture is summarized in Fig.~\ref{fig:fig2}. In Fig.~\ref{fig:fig2} a, the current noisy state $x_t$ from the forward process and the diffusion timestep $t$ are encoded together with a spatiotemporal representation of the multi-round syndrome history. This syndrome-conditioned representation is then further adjusted through the modulation mechanism shown in Fig.~\ref{fig:fig2} b. Given $(x_t,t,s)$, the denoiser produces bitwise logits corresponding to a clean-state distribution$p_\theta(x_0 \mid x_t, s)$,
where $s$ denotes the measured syndrome history.

For the intermediate reverse steps, the predicted clean-state distribution is combined with the discrete forward kernels to construct a reverse posterior over $x_{t-1}$. Specifically, the reverse update uses the one-step transition kernel $Q_t$ and the cumulative kernel $\bar Q_{t-1}$, and marginalizes over the predicted clean state:
\[
p_\theta(x_{t-1}\mid x_t,s)
=
\sum_{x_0}
q(x_{t-1}\mid x_t,x_0)\,p_\theta(x_0\mid x_t,s),
\]
where \(q(x_{t-1}\mid x_t,x_0)\) is the posterior transition induced by the discrete forward process. Since the binary corruption process factorizes over coordinates, this posterior is evaluated bitwise, and the next state $x_{t-1}$ is sampled independently for each bit. This reverse-posterior construction follows the general formulation of discrete-state diffusion models~\cite{austin2021structured}.

After iterating these posterior-guided denoising steps from $t=T$ down to $1$, the final decoded output is obtained at $t=0$ by taking the bitwise argmax of the denoiser logits. The entire reverse process therefore remains in the binary domain $\{0,1\}^{L}$, ensuring that decoding is performed directly in the discrete logical space.

\subsection*{Experimental settings}
In this section, we summarize the experimental settings of our paper.
We evaluate DiffQEC on the publicly released Google Sycamore memory 
dataset~\cite{Bausch2024} for rotated surface codes with distances 
$d = 3$ and $d = 5$, which contains both X-basis and Z-basis memory experiments. 
For $d = 3$, experiments are performed at four chip locations; for $d = 5$ at a 
single location. In the single-round setting, the decoder takes the syndrome from 
one error-correction round as input and predicts the final logical error outcome 
$l \in \{0,1\}^L$. To prevent data leakage between training and evaluation, we 
employ a parity-based data split: DEMs are fitted separately on even- and 
odd-indexed shots, and models are trained on DEM-generated samples from one 
parity subset and evaluated exclusively on the held-out experimental shots of the 
other parity. In the multi-round setting, the decoder takes the full syndrome 
history across $r \in \{3, 5, 7, 9\}$ rounds as input; training again relies on 
DEM-generated samples rather than experimental shots directly. Rather than 
training a separate model for each round count, we jointly train a single model 
across all values of $r$ using a curriculum strategy: training begins by cycling 
through shorter syndrome histories before progressively introducing longer ones, 
encouraging the model to first learn temporally local correlations before 
handling longer-range temporal dependencies.
To assess scalability beyond current experimental sizes, we study simulated 
surface-code memory experiments under the standard depolarizing noise model at 
code distances up to $d = 17$.

For circuit-level evaluation, we consider mirror-symmetric random logical circuits. Specifically, a sequence of random logical gates is applied in the first half of the circuit and then reversed in the second half. In the absence of noise, this construction returns the logical state to its initial value, so that any mismatch between the final logical measurement and the initial logical preparation can be attributed to errors accumulated during circuit execution. We simulate these circuits under circuit-level noise on rotated surface-code logical qubits with code distance 3. Decoder inputs are the full syndrome histories obtained from repeated stabilizer measurements during circuit execution, and the prediction target is the final logical error outcome. We evaluate decoder performance across logical circuit depths from 1 to 5.

\noindent \textbf{Training objective}. Training is performed by randomly sampling a diffusion step $t$, generating a corrupted state $x_t \sim q(x_t\mid x_0)$ from the clean target $x_0$, and asking the denoiser to recover $x_0$ from $(x_t,t,s)$. Let $z_\theta(x_t,t,s)\in\mathbb{R}^{L\times 2}$ denote the output logits. We optimize the bitwise cross-entropy loss
\[
\mathcal{L}=
\mathbb{E}_{x_0,s,t,x_t}
\left[
\frac{1}{L}\sum_{i=1}^{L}
\mathrm{CE}\!\left(z_\theta^{(i)}(x_t,t,s),\,x_0^{(i)}\right)
\right],
\]
where $x_0^{(i)}$ is the $i$th binary target and $\mathrm{CE}$ denotes the two-class cross-entropy loss for that bit.

\subsection*{Network Architecture}

The overall architecture of DiffQEC is illustrated in Fig.~\ref{fig:fig2}. 
The denoising network takes three inputs: the multi-round syndrome history $s$, 
the current noisy state $x_t$, and the diffusion timestep $t$. These are first 
encoded separately and then combined to form a joint representation that is 
iteratively refined through a stack of syndrome-conditioned denoising layers, 
producing bitwise logits over the logical correction $\hat{x}_0$.

\noindent\textbf{Syndrome processor.}
As shown in Fig.~\ref{fig:fig2}a, the syndrome history is encoded through a 
two-stage spatial-then-temporal pipeline. In the spatial stage, each round's 
syndrome bits are arranged into a $d \times d$ spatial map and processed by a 
shared convolutional network followed by pooling, yielding a per-round feature 
vector. These per-round features are then aggregated temporally by a gated 
recurrent unit (GRU)~\cite{cho-etal-2014-learning}, whose final hidden state 
serves as the syndrome feature $\mathbf{c}$. In parallel, the diffusion timestep 
$t$ and the noisy state $x_t$ are encoded separately via a time encoder and an 
MLP respectively. The three representations are combined by addition to form the 
encoded feature,
$\mathbf{h}_0 = \mathbf{e}_x + \mathbf{e}_t + \mathbf{c}$,
which is then passed to the denoising network.

\noindent\textbf{Syndrome feature modulation.}
As shown in Fig.~\ref{fig:fig2}b, each denoising layer further conditions its 
computation on the syndrome feature $\mathbf{c}$ through a syndrome controller 
that generates scale and shift signals to modulate the intermediate 
representation. The modulated output is then combined with the layer input 
through a gate residual pathway, which allows each layer to selectively 
incorporate the syndrome-conditioned update while maintaining stable gradient 
flow. This design keeps the denoising computation persistently conditioned on 
the observed syndrome throughout the network. The final layer produces bitwise 
logits $z_\theta(x_t, t, s) \in \mathbb{R}^{L \times 2}$ over the logical 
correction $\hat{x}_0$.

\bibliography{sn-bibliography}
\appendix
\end{document}